\documentclass[aip,graphicx]{revtex4-1}
\usepackage[cp1251]{inputenc}
\usepackage[T2A]{fontenc}
\usepackage[english]{babel}
\usepackage{amssymb,latexsym,amsmath,amscd}
\usepackage{graphicx,color,framed}
\begin{document}
\selectlanguage{english}

\title{A Statistical Theory of cosolvent-induced coil-globule transitions in dilute polymer solution}

\author{\firstname{Yu.~А.} \surname{Budkov}}
\email[]{urabudkov@rambler.ru}
\affiliation{Institute of Solution Chemistry of the Russian Academy of Sciences, Ivanovo, Russia}

\author{\firstname{ A.~L.} \surname{Kolesnikov}}
\affiliation{Ivanovo State University, Ivanovo, Russia}
\affiliation{Institut f\"{u}r Nichtklassische Chemie e.V., Universitat Leipzig,
Leipzig, Germany}
\author{\firstname{ N.} \surname{Georgi}}
\affiliation{Max Planck Institute for Mathematics in the Sciences, Leipzig, Germany}

\author{\firstname{ M.~G.} \surname{Kiselev}}
\affiliation{Institute of Solution Chemistry of the Russian Academy of Sciences, Ivanovo, Russia}
\begin{abstract}
We present a statistical model of a dilute polymer solution in good solvent in the presence of low-molecular weight cosolvent. 
We investigate the conformational changes of the polymer induced by a change of the cosolvent concentration and the type of 
interaction between the cosolvent and the polymer. We describe the polymer in solution by the Edwards model, 
where the partition function of the polymer chain with a fixed radius of gyration is described in the 
framework of the mean-field approximation. The contributions of polymer-cosolvent and the cosolvent-cosolvent 
interactions in the total Helmholtz free energy are treated also within the mean-field approximation. 
For convenience we separate the system volume on two parts: the volume occupied by the polymer chain 
expressed through its gyration volume and the bulk solution. Considering the equilibrium between the 
two subvolumes we obtain the total Helmholtz free energy of the solution as a function of radius 
of gyration and the cosolvent concentration within gyration volume.
After minimization of the total Helmholtz free energy with respect to its arguments
we obtain a system of coupled equations with respect to the radius of gyration of the polymer chain 
and the cosolvent concentration within the gyration volume. Varying the interaction strength between polymer 
and cosolvent we show that the polymer collapse occurs in two cases - either when the interaction 
between polymer and cosolvent is repulsive or when the interaction is attractive. The reported
effects could be relevant for different disciplines where conformational transitions of macromolecules 
in the presence of a cosolvent are of interest, in particular in biology, chemistry and material science. 
\end{abstract}

\maketitle
\section{Introduction}
The coil-globule transition in dilute polymer solutions is one of the most fascinating phenomena
in physical chemistry of polymers. The mechanism of conformational transition of a chain molecule upon a change of 
environment has found many applications in recent technological advances ranging from encapsulation of drug molecules 
in a polymer coil and targeted delivery  \cite{DrugDeliveryReview,DrugDeliveryReview2,DrugDeliveryReview3} 
to smart materials changing their properties in response to the environment \cite{SmartBioPolyReview,Smartpolymers}. 
Numerous applications are based on a conformational transition of a polymer sensing the presence of specific molecules at low concentrations \cite{SensorsBook}, 
inducing a phase change aggregation of suspensions of colloids coated with pH or temperature responsive polymers \cite{SmartHydrogelBook},  
gels comprising thermoresponsive polymer networks \cite{SmartHydrogelBook,SmartColloidsBook} to name only a few. 
In organisms proteins fold into a compact state attaining a well defined biological function by exposing functional groups to their environment. 
Viral DNA collapses to a condensed state to fit in the confined space of a viral capsid \cite{CondensedDNAGelbart}. 
First steps in DNA separation for subsequent analysis involve condensation of DNA using osmolytes or denaturants 
\cite{DNAInteractions,DNAInteractionsColloidsInterfaces}. 
Especially water soluble polymers are used to exert a lateral pressure on 
the DNA to induce a collapse \cite{CondensedDNAReview}. 
In the present article an alternative mechanism is outlined that could also lead to a collapse when the low 
molecular weight cosolvent has entered the coil of a chain molecule compressing the coil from within. 

The ubiquitous presence of chain molecules and the principal possibility to control the conformational 
transition by an external stimulus has therefore attracted much attention from experimental point of view in chemistry, 
biology and material science.

Theoretical efforts on the other hand formulating a coil-globule transition theory 
contributed much to a qualitative understanding of this phenomenon. Today many theoretical 
contributions exist dedicated to the coil-globule transition ranging from simplified self-consistent 
field treatment of the solvent to theories based on the field-theoretic formalism 
\cite{Lifshitz,Lifshitz_Grosberg,Moore,deGennes_collaps,Birshtein_1,Birshtein_2,Kholodenko,Khohlov,Khohlov_2,Muthukumar,Sanchez,
Wu_PRL,Sear}. It has been shown that, as the solvent becomes poorer, 
the polymer coil shrinks leading eventually to a collapse of the polymer coil. 
Predominantly the theoretical models describe the solvent only implicitly, 
i.e. its influence on macromolecule taken into account through effective monomer-monomer interaction. 
However, nowadays, due to the importance of conformation control of macromolecules in solution by low-molecular weight
cosolvents (for example, adding of osmolytes or denaturants in protein aqueous solutions 
\cite{JACS_2004,JACS_2010,Cosolvent_2013,Protein_collaps_2012,Collaps_2009}) 
explicit account of the cosolvent in the model seems to be indispensable.

However, up to now only a few attempts considering the cosolvent explicitly have been reported. 
Notably in the work of Tanaka, et al. \cite{Tanaka}  the conformational phase transition 
of an isolated polymer chain capable of forming physical bonds with solvent molecules treated the solvent explicitly. 
On the basis of a Flory type mean-field theory, a formula for the temperature dependence 
of the expansion parameter of the chain has been derived. The formation of the physical bonds 
between polymer and solvent molecules causes a reentrant conformational change between 
coiled and globular state of the polymer chain when temperature is varied.

In the work \cite{Collaps_2013} the collapse and swelling behavior of a homopolymer 
has been studied using implicit-solvent, explicit cosolvent Langevin dynamics computer simulations. 
Varying the interaction strength the results of two theoretical models have been compared 
with the simulation findings. The first model was based on an effective one-component description where the 
co-solutes have been averaged out - the second model treated the fully two-component system 
in a Flory-de Gennes type of approach. A conclusion has been reached that the simulation results were 
in accord with the predictions of the second model.

However, to our best knowledge, there does not seem to exist an approach starting 
from first principals of statistical mechanics describing the influence of the cosolvent 
on conformational behavior of the polymer chain -- specifically the concentration dependence of the radius 
of gyration of the polymer chain taking into account the type of cosolvent interactions with the polymer. 
In the present work such a statistical model of a flexible polymer chain in good solvent in the presence 
of a low-molecular cosolvent is developed. The influence of cosolvent concentration and
quality of its interaction with monomers on the collapse and swelling behavior of polymer chain is investigated.

The paper organized as follows. In Sec. II, we present our theoretical model and in Sec. III
the numerical results are given. In Sec. IV, we discuss the obtained results and summarize our findings.

\section{Theory}
We consider the case of a polymer chain immersed in a good solvent. We will describe the polymer 
in the framework of the Edwards model (flexible polymer chain with excluded volume) \cite{Edwards_1,Edwards_2}. 
The polymer chain molecule and the low-molecular weight cosolvent at a specified 
concentration are immersed in a solvent described by a continuous, structureless medium.
We assume further that the volume of the system consists of two parts: the gyration 
volume containing predominantly monomers of the polymer chain and a bulk solution (see, fig. 1). 
Thus, we consider the cosolvent concentration at equilibrium  in the two subvolumes varying the strength 
of interaction of the polymer-cosolvent. In order to find the solution of the posed problem 
the minimum of total Helmholtz free energy of the system as function of the radius of gyration 
and the number of cosolvent molecules within gyration volume is sought.

We assume further that the interaction between cosolvent molecules is purely repulsive so that it can be described by 
simple hard-sphere potential. This assumption is reasonable if a thermodynamic state of cosolvent 
is far from binodal in region above the critical point or when it is in the field of liquid states 
where attraction between cosolvent molecules is not important.  Basically, we consider 
the theory at the level of the mean-field approximation. 
However, we take into account the cosolvent density fluctuation effect on the monomer-monomer interaction.
Our aim is to study the dependence of polymer chain conformations as a function 
of the cosolvent concentration and the type of interaction between cosolvent and  monomers.

\begin{figure}[h]
\center{\includegraphics[width=0.5\linewidth]{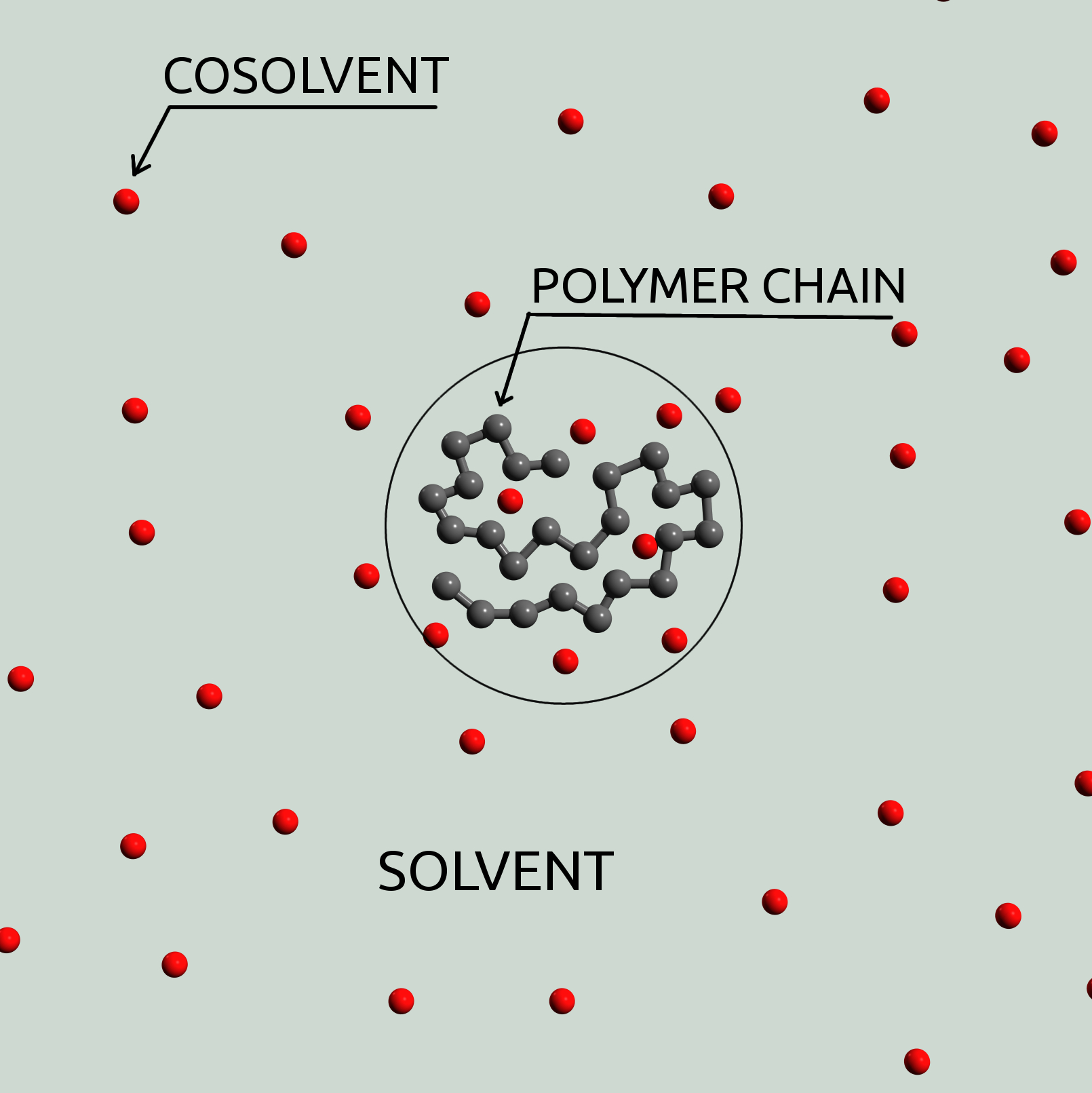}}
\caption{\sl Pictorial representation of the model. The system consists of a polymer 
chain in a good solvent which is represented by a continuous, structureless medium 
and a low-molecular weight cosolvent at a specified concentration. The total volume 
of the system consists of two parts: the gyration volume containing predominantly monomers 
of the polymer chain and a bulk solution.}
\label{Fig:1}
\end{figure}

The partition function of the polymer-cosolvent system takes the form (for details, see Appendix)
\begin{equation} \label{PartSum:Total}
Z(R_{g},N_{1})=\mathcal{Z}_{c}(R_{g},N_{1})Z_{p}(R_{g}),
\end{equation}
where 
\begin{equation} 
\label{PartSum:CoSolv}
\mathcal{Z}_{c}(R_{g},N_{1})=\frac{\left(V_{g}-N_{1}v_{c}-Nv_{mc}\right)^{N_{1}}(V-V_{g}-(N_{c}-N_{1})v_{c})^{N_{c}-N_{1}}}{(N_{c}-N_{1})!N_{1}!}e^{-\beta\Delta F_{int}(N_{1})}
\end{equation}
is a partition function of low-molecular weight cosolvent in  solution;
\begin{equation}
\label{Fluct}
\beta \Delta F_{int}(N_{1})=\frac{w_{pc}NN_{1}}{V_{g}-Nv_{mc}}-\frac{w_{pc}^2N_{1}\chi_{T}}{2\chi_{T}^{id}}\frac{N^2}{V_{g}^2},
\end{equation}
$V$ is a volume of the system; $N_{1}$ is a number of 
cosolvent molecules which are within the gyration volume 
$V_{g}=\frac{4\pi  R_{g}^3}{3}$ ($R_{g}$ is the radius of gyration); 
$N$ is a degree of polymerization - e.g. length of the polymer chain; $w_{pc}$ 
is a parameter of volume interaction polymer-cosolvent corresponding to the second virial coefficient; 
$N_{c}$ is a total number of cosolvent molecules, $v_{c}=\frac{\pi d_{c}^3}{6}$ ($d_{c}$ 
is cosolvent molecules diameter) is the excluded volume of cosolvent molecules, 
$v_{mc}=\frac{\pi d_{mc}^3}{6}$ ($d_{mc}=\frac{d_{m}+d_{c}}{2}$), $d_{m}$ is a diameter of monomers; 
$\chi_{T}$ is an isothermal compressibility of cosolvent in the gyration volume and $\chi_{T}^{id}$ 
is an isothermal compressibility of ideal gas. The expression for the isothermal compressibility 
in the case of pure repulsive interaction within the mean-field approximation has a form 
\begin{equation}
\label{compressibility}
\chi_{T}(\rho,T)=\frac{1}{\rho}\left(\frac{\partial \rho}{\partial P}\right)_{T}=\frac{(1-\rho v_{c})^2}{\rho k_{B}T},
\end{equation}
where $\rho$ is a cosolvent concentration, $k_{B}$ is a Boltzmann constant, $T$ is a temperature, $P=\frac{\rho k_{B}T}{1-\rho v_{c}}$ is a
pressure. The first term in the right hand side of (\ref{Fluct}) is 
relates to the monomer-cosolvent interaction in the framework of mean-field approximation. 
The second term in (\ref{Fluct}) is a contribution of the effective monomer-monomer interaction 
due to cosolvent concentration fluctuations.

The second term in the product (\ref{PartSum:Total})
\begin{equation} \label{PartSum:Poly1}
Z_{p}(R_{g})=\int d\Gamma (R_{g})e^{-\beta H_{p}}
\end{equation}
is a partition function of the polymer chain with fixed radius of gyration; the symbol $\int d\Gamma(R_{g})(..)$ 
denotes an integration over microstates of the polymer chain 
at a fixed radius of gyration; $H_{p}$ is a Hamiltonian of interaction monomer-monomer
which includes the hard-core potential and simple 1-body monomer 
interaction potential due to the solvent influence (see Appendix).
Using the mean-field approximation we obtain the following 
expression for the polymer Helmholtz free energy
\begin{equation} \label{FreeEner}
\beta F_{p}(R_{g})=-\ln{Z_{p}(R_{g})}=\beta F_{id}(R_{g})+\frac{N^2 w_{p}}{2 V_{g}}-N\ln\left(1-\frac{Nv_{m}}{V_{g}}\right),
\end{equation} \label{FreeEnerId}
where $\beta F_{id}(R_{g})=-\ln Z_{id}(R_{g})$ is the Helmholtz free energy of the ideal polymer chain, 
$\beta =\frac{1}{k_{B}T}$, $v_{m}=\frac{\pi d_{m}^3}{6}$ is a monomer excluded volume. 
The second term in (\ref{FreeEner}) determines the contribution of monomer-monomer interaction due to solvent effect.
The last term in (\ref{FreeEner}) corresponds to the contribution of hard-sphere effect in the
Helmholtz polymer free energy within the mean-field approximation. 
Based on the results of Fixman \cite{Fixman_Gyration} we construct an interpolation formula 
for the free energy of the ideal polymer chain:
\begin{equation} \label{FreeEnerInter}
\beta F_{id}(R_{g})=\frac{9}{4}\left(\alpha^2+\frac{1}{\alpha^2}\right)-\frac{3}{2}\ln{\alpha^2},
\end{equation}
where $\alpha=\frac{R_{g}}{R_{0g}}$ denotes the expansion parameter, $R_{0g}^2=\frac{Nb^2}{6}$ 
is a mean-square radius of gyration of the ideal polymer chain and $b$ is the Kuhn length of the segment.
Rewriting the polymer free energy in terms of the expansion parameter $\alpha$ we obtain
\begin{equation} \label{FreeEnerInterFinal}
\beta F_{p}(\alpha)=\frac{9}{4}\left(\alpha^2+\frac{1}{\alpha^2}\right)-\frac{3}{2}\ln{\alpha^2}+
\frac{9\sqrt{6}w_{p}\sqrt{N}}{4\pi b^3\alpha^3}-N\ln\left(1-\frac{9\sqrt{6}v_{m}}{2\pi\sqrt{N}\alpha^3b^3}\right).
\end{equation}
The expression for the cosolvent Helmholtz free energy takes the form
\begin{equation} \label{FreeEnerCosolvent1}
\beta F_{c}(R_{g},N_{1})=-\ln{\mathcal{Z}_{c}(R_{g},N_{1})}=\beta\Delta F_{int}(N_{1})-N_{1}\ln{\left(V_{g}-N_{1}v_{c}-Nv_{mc}\right)}-\nonumber
\end{equation}
\begin{equation}\label{FreeEnerCosolvent2}
-\left(N_{c}-N_{1}\right)\ln(V-V_{g}-(N_{c}-N_{1})v_{c})+N_{1}\left(\ln{N_{1}}-1\right)+(N_{c}-N_{1})\left(\ln(N_{c}-N_{1})-1\right).
\end{equation}
Minimizing $\beta F_{c}(R_{g},N_{1})$ with respect to $N_{1}$, i.e. equating to zero the derivative
$\frac{\partial{\left(\beta F_{c}(R_{g},N_{1})\right)}}{\partial{N}_{1}}$ and introducing the notations 
$\rho_{1}=\frac{N_{1}}{V_{g}}$ and $\rho=\frac{N_{c}}{V}$
we obtain the equation for the density of the cosolvent within the gyration volume $\rho_{1}$
\begin{equation} \label{DensFinal}
\frac{\rho_{1}}{1-\rho_{1}v_{c}-\frac{9\sqrt{6}v_{mc}}{2\pi\sqrt{N}\alpha^3 b^3}}=
\end{equation}
\begin{equation}
=\frac{\rho}{1-\rho v_{c}}e^{-\frac{\rho_{1}v_{c}}{1-\rho_{1}v_{c}-\frac{9\sqrt{6}v_{mc}}{2\pi\sqrt{N}\alpha^3b^3}}+\frac{\rho v_{c}}{1-\rho v_{c}}-
\frac{9\sqrt{6}w_{pc}}{2\pi \sqrt{N}\alpha^3 b^3\left(1-\frac{9\sqrt{6}v_{mc}}{2\pi\sqrt{N}\alpha^3 b^3}\right)}+\frac{243w_{pc}^2}{4\pi^2N\alpha^6b^6}A(\rho_{1})},
\end{equation}
which valid for $V\gg V_{g}$ and $N_{c}\gg N_{1}$; $A(\rho_{1})=\left(1-\rho_{1}v_{c}\right)\left(1-3\rho_{1}v_{c}\right)$.

It should be noted that the value of the expansion parameter, which corresponds to a minimum of the total Helmholtz free energy.
Thus, using the equations (\ref{compressibility},\ref{FreeEnerInterFinal}-\ref{DensFinal}), and calculating a derivative of the total free energy 
with respect to $\alpha$ and equating it to zero, we obtain the equation
\begin{equation} \label{AlphaPolynom}
\alpha^5-\frac{2}{3}\alpha^3-\alpha=\frac{3\sqrt{6}}{2\pi}\sqrt{N}\left(\tilde{w_{p}}+\frac{\tilde{v}_{m}}{1-\frac{9\sqrt{6}\tilde{v}_{m}}{2\pi \sqrt{N}\alpha^3}}-B(\tilde{\rho}_{1})\right)-\nonumber
\end{equation}
\begin{equation}
-\frac{2\pi\sqrt{6}}{81}N^{3/2}\alpha^6\left(\frac{\tilde{\rho}}{1-\tilde{\rho}\tilde{v}_{c}}-\frac{\tilde{\rho}_{1}}{1-\tilde{\rho}_{1}\tilde{v}_{c}-\frac{9\sqrt{6}\tilde{v}_{mc}}{2\pi\sqrt{N}\alpha^3}}\right)+\frac{2N\tilde{w}_{pc}\tilde{\rho}_{1}\alpha^3}{3\left(1-\frac{9\sqrt{6}\tilde{v}_{mc}}{2\pi\sqrt{N}\alpha^3}\right)},
\end{equation}
where $\tilde{w}_{p}=w_{p}b^{-3}$, $\tilde{w}_{pc}=w_{pc}b^{-3}$, $\tilde{\rho}=\rho b^3$, $\tilde{v}_{c}=\frac{v_{c}}{b^3}$, 
$\tilde{v}_{mc}=\frac{v_{mc}}{b^3}$, $\tilde{v}_{m}=\frac{v_{m}}{b^3}$, $B(\tilde{\rho}_{1})=2\tilde{w}_{pc}^2\tilde{\rho}_{1}\left(1-2\tilde{\rho}_{1}\tilde{v}_{c}\right)\left(1-\tilde{\rho}_{1}\tilde{v}_{c}\right)$.
The first term in a right hand side of equation (\ref{AlphaPolynom}) relates to the monomer-monomer 
interaction due to solvent effect, monomer hard-core effect and effect of cosolvent concentration fluctuations. 
The second term relates to a pressure difference between the cosolvent molecules within gyration 
volume and in the bulk solution. The last term is related to the polymer-cosolvent interaction.

The dimensionless cosolvent concentration $\tilde{\rho}_{1}=\rho_{1}b^3$ within gyration volume satisfies the equation
\begin{equation}  
\frac{\tilde{\rho}_{1}}{1-\tilde{\rho}_{1}\tilde{v}_{c}-\frac{9\sqrt{6}\tilde{v}_{mc}}{2\pi\sqrt{N}\alpha^3}}=\nonumber
\end{equation}
\begin{equation}
\label{AlphaPolynom2}
=\frac{\tilde{\rho}}{1-\tilde{\rho}\tilde{v}_{c}}
e^{-\frac{\tilde{\rho}_{1}\tilde{v}_{c}}{1-\tilde{\rho}_{1}\tilde{v}_{c}-\frac{9\sqrt{6}\tilde{v}_{mc}}{2\pi\sqrt{N}\alpha^3}}+\frac{\tilde{\rho}\tilde{v}_{c}}{1-\tilde{\rho}\tilde{v}_{c}}-
\frac{9\sqrt{6}\tilde{w}_{pc}}{2\pi \sqrt{N}\alpha^3\left(1-\frac{9\sqrt{6}\tilde{v}_{mc}}{2\pi\sqrt{N}\alpha^3}\right)}+\frac{243\tilde{w}_{pc}^2}{4\pi^2N\alpha^6}A(\tilde{\rho_{1}})}.
\end{equation}

Let us analyze some limiting regimes. At $\tilde\rho\rightarrow 0$ we have a 
swelling regime $\alpha \sim (\tilde{w}_{p}+\tilde{v}_{m})^{1/5}N^{1/10}$ 
that is described by the classical mean-field Flory theory \cite{Flory_book}. Now, 
we consider the situation when $\tilde{w}_{pc}\gg 1$, i.e. when 
interaction cosolvent-polymer is strongly repulsive. In this case $\tilde{\rho}_{1}\ll \tilde{\rho}$ 
and the equation (\ref{AlphaPolynom}) simplifies to 
\begin{equation} \label{AlphaPolynomSimple} 
\alpha^5-\frac{2}{3}\alpha^3-\alpha=\frac{3\sqrt{6}}{2\pi}\sqrt{N}\frac{\tilde{v}_{m}}{1-\frac{9\sqrt{6}\tilde{v}_{m}}{2\pi \sqrt{N}\alpha^3}}-\frac{2\pi\sqrt{6}}{81}\frac{\tilde{\rho}N^{3/2}\alpha^6}{1-\tilde{\rho}\tilde{v}_{c}}.
\end{equation}
If the second term on the right hand side of equation (\ref{AlphaPolynomSimple}) dominates
then neglecting all except the first and second terms we obtain the following relations 
for expansion parameter and radius of gyration
\begin{equation} \label{AlphaRg} 
\alpha \simeq a^{1/3}N^{-\frac{1}{6}},~~~ 
\frac{R_{g}}{b}\simeq \frac{\sqrt{6}}{6}a^{1/3}N^{1/3},
\end{equation}
which corresponds to a globular conformation; 
$a=\frac{9\sqrt{6}\tilde{v}_{m}}{4\pi}+\sqrt{\left(\frac{9\sqrt{6}\tilde{v}_{m}}{4\pi}\right)^2+\frac{243\tilde{v}_{m}(1-\tilde{\rho}\tilde{v}_{c})}{4\pi^2\rho}}$.

We turn now to the opposite limiting case when  $\tilde{w}_{pc}< 0$ and $|\tilde{w}_{pc}|\gg 1$, i.e. when interaction 
cosolvent-polymer is strongly attractive. In this case $\tilde{\rho}_{1}\gg  \tilde{\rho}$. Therefore the difference 
of cosolvent pressures between interior of the gyration volume and the bulk can lead to an additional swelling the polymer coil. 
The excluded volume of cosolvent molecules and monomers has the same affect. 
However, the strong attraction between cosolvent and monomers leads to a shrinking of the 
polymer coil. Due to the competition between these trends the coil-globule transition can occur. 
The equation (\ref{AlphaPolynom}) in this case simplifies to
\begin{equation}
\label{globule_size}
\frac{2\sqrt{N}|\tilde{w}_{pc}|\tilde{\rho}_{1}\alpha^3}{3\left(1-\frac{9\sqrt{6}\tilde{v}_{mc}}{2\pi\sqrt{N}\alpha^3}\right)}\simeq \frac{2\pi\sqrt{6}N\alpha^{6}}{81}\frac{\tilde{\rho_{1}}}{1-\tilde{\rho}_{1}\tilde{v}_{c}-\frac{9\sqrt{6}\tilde{v}_{mc}}{2\pi\sqrt{N}\alpha^3}}
+\frac{3\sqrt{6}}{2\pi}\frac{\tilde{v}_{m}}{1-\frac{9\sqrt{6}\tilde{v}_{m}}{2\pi\sqrt{N}\alpha^3}}.
\end{equation}
The equation (\ref{globule_size}) can be simplified by the substitution $\alpha = s N^{-1/6}$. 
Thus we obtain the equation with respect to $s$ which already does not contain the degree of polymerization $N$
\begin{equation}
\label{globule_size_2}
\frac{\pi\sqrt{6}}{27}\frac{s^{6}\tilde{\rho}_{1}}{1-\tilde{\rho}_{1}\tilde{v}_{c}-\frac{9\sqrt{6}\tilde{v}_{mc}}{2\pi s^3}}+\frac{9\sqrt{6}}{4\pi}\frac{\tilde{v}_{m}}{1-\frac{9\sqrt{6}\tilde{v}_{m}}{2\pi s^3}}-\frac{|\tilde{w}_{pc}|\tilde{\rho}_{1}s^3}{1-\frac{9\sqrt{6}\tilde{v}_{mc}}{2\pi s^3}}=0.
\end{equation}
Therefore, the equation (\ref{globule_size_2}) determines the globule size as function of $\tilde{v}_{c}$, $\tilde{v}_{m}$, $\tilde{v}_{mc}$ and $\tilde{w}_{pc}$: 
\begin{equation} 
\label{AlphaRepulsion}
\alpha\simeq sN^{-1/6}, ~~ \frac{R_{g}}{b}\simeq \frac{\sqrt{6}}{6}sN^{1/3}.
\end{equation}
In this case size of the globule is determined by a competition between cosolvent-polymer attraction 
and monomer and cosolvent excluded volume effect. We would like to stress that in the globular regimes 
the cosolvent concentration $\tilde{\rho}_{1}$ within gyration volume does not depend on the expansion parameter.

\section{Numerical results and discussion}
Turning to the numerical analysis of the system of equations (\ref{AlphaPolynom}-\ref{AlphaPolynom2}) 
we will fix the diameters (in units of Kuhn's length of the segment $b$) $\tilde{d}_{c}=\tilde{d}_{m}=\tilde{d}_{mc}=1$, 
parameter of volume interactions monomer-monomer $\tilde{w}_{p}=1$, and degree of polymerization $N=10^{4}$.

We first discuss the case when the interaction polymer-cosolvent is a pure repulsive.
Fig. 2 (a) shows the expansion parameter $\alpha$ as a function of the cosolvent 
concentration $\tilde{\rho}$ at different values of $\tilde{w}_{pc}$. 
At increasing  cosolvent concentration the expansion parameter monotonically 
decreases and is asymptotically close to limit given by (\ref{AlphaRg}) 
corresponding to a globular conformation. Increasing the interaction parameter 
$\tilde{w}_{pc}$ the coil-globule transition becomes sharper.
Fig. 2 (b) shows the cosolvent concentration in the gyration volume as 
a function of cosolvent concentration in the bulk for two values of 
polymer-cosolvent interaction parameters.
In the region of coil-globule transition the cosolvent concentration 
within the gyration volume shows a sufficiently pronounced maximum. 
Such behavior can be interpreted as follows. 
At small values of cosolvent concentrations in the bulk the gyration volume 
offers enough space for the cosolvent to intrude leading to a swelling of the polymer coil.
In contrast, increasing the size of polymer coil decreases leading to a more confined space increasing therefore 
the repulsion between cosolvent and monomers and as consequence the cosolvent is expelled from polymer coil.
Thus, when the cosolvent concentration in the bulk drops below a certain value the cosolvent 
concentration within the gyration volume monotonically decreases.

\begin{figure}[h]
\center{\includegraphics[width=0.65\linewidth]{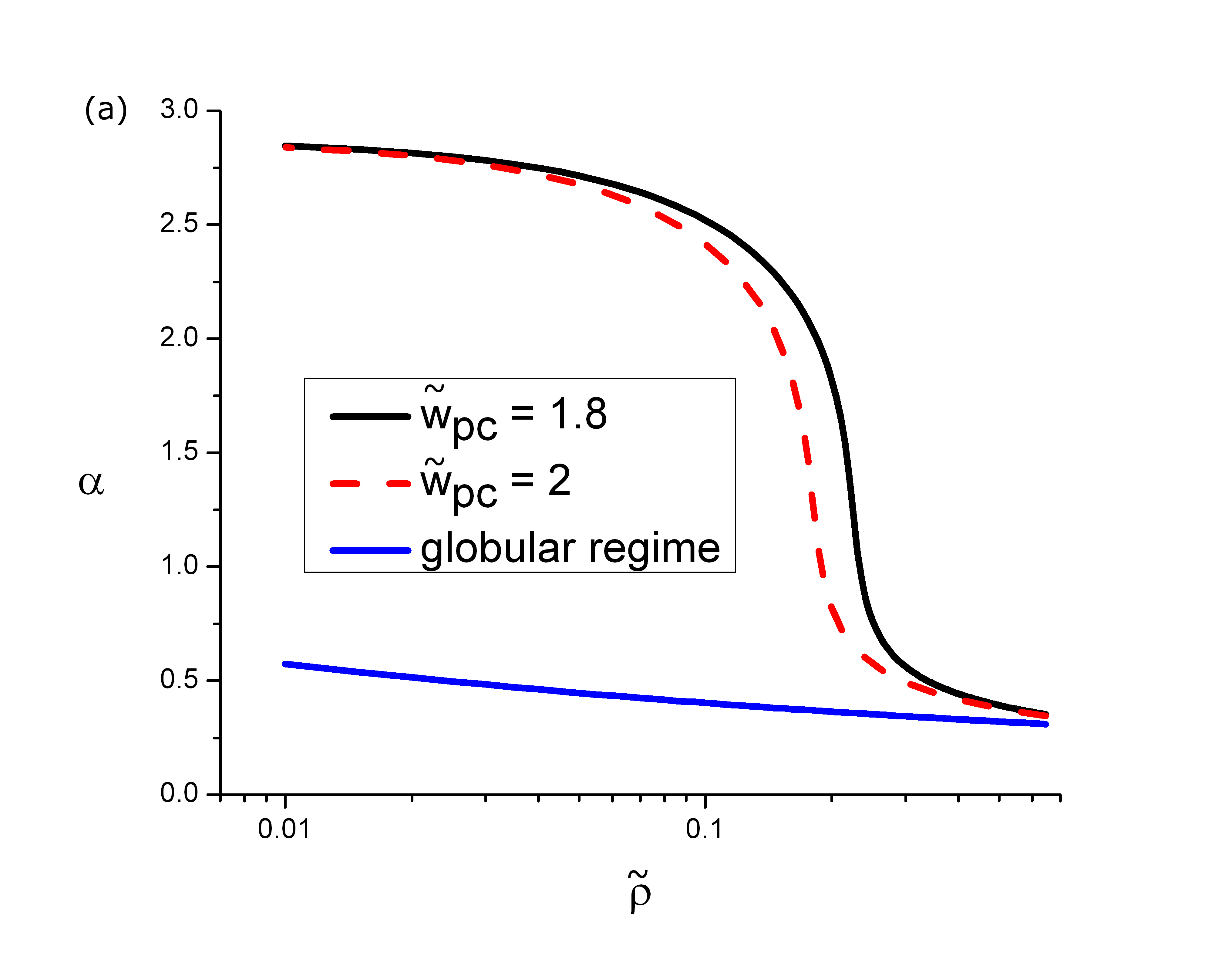}}
\center{\includegraphics[width=0.65\linewidth]{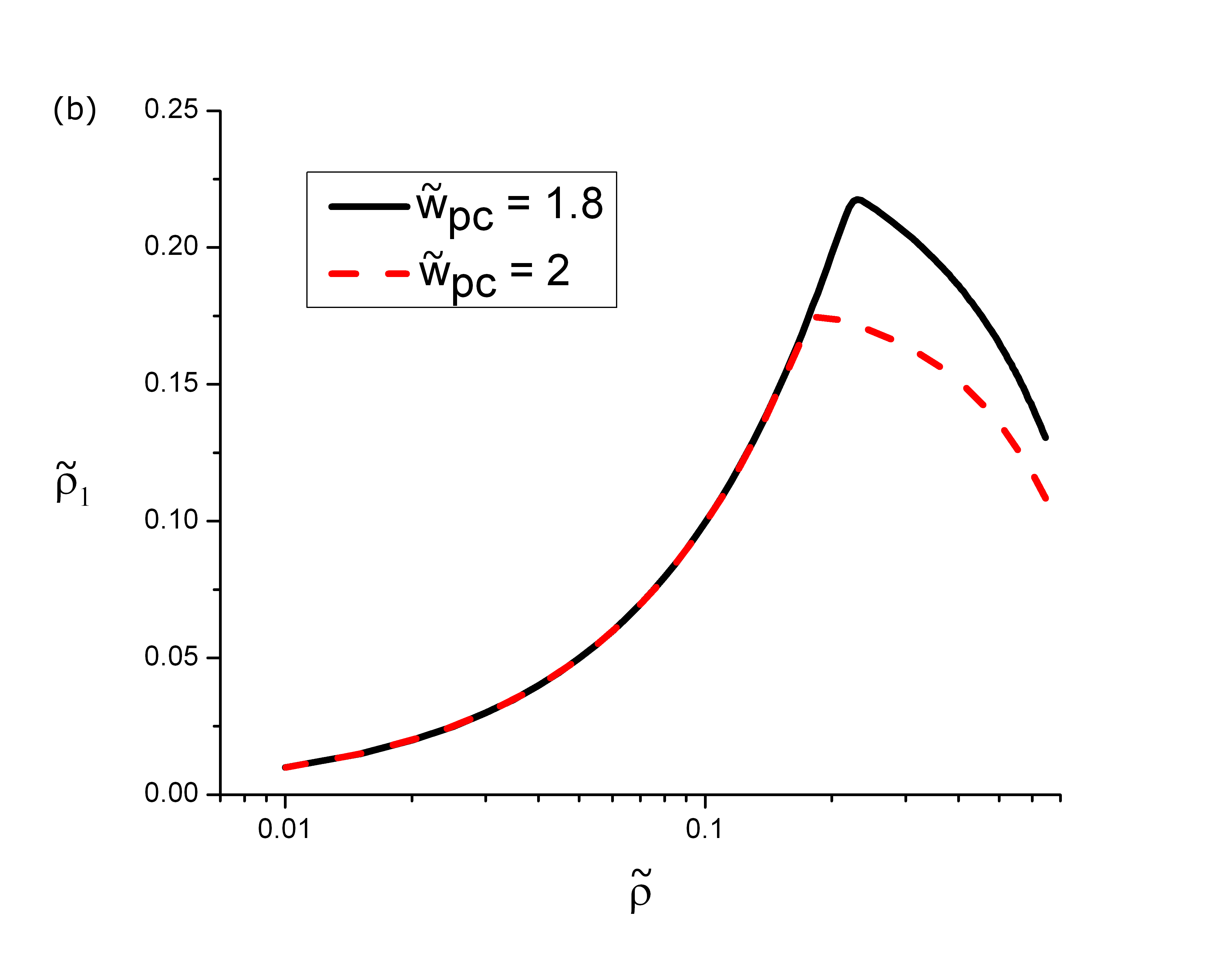}}
\caption{\sl  Repulsive polymer-cosolvent interaction ( e.g. $\tilde{w}_{pc}>0$ ). 
(a) The expansion parameter $\alpha$ as  a function of cosolvent 
concentration $\tilde{\rho}$ in the bulk solution. (b) The cosolvent concentration 
in gyration volume $\tilde{\rho}_{1}$ as a function on the cosolvent bulk concentration $\tilde{\rho}$ at different 
positive parameter of interaction polymer-cosolvent $\tilde{w}_{pc}$.
It is seen that at increasing of cosolvent concentration the expansion parameter 
is monotonically decrease and asymptotically close to limit 
(25) corresponding to globular conformation.
It is easy to seen that at increasing of parameter $\tilde{w}_{pc}$ 
the coil-globule transition becomes sharper. 
In the region of coil-globule transition the cosolvent concentration 
within the gyration volume shows a sufficiently pronounced maximum. 
We use $\tilde{w}_{p}=1$, $N=10^4$, $\tilde{d}_{m}=\tilde{d}_{c}=\tilde{d}_{mc}=1$.}
\label{Fig:2}
\end{figure}

For the case when the polymer-cosolvent interaction is attractive ($\tilde{w}_{pc}<0$) 
an abrupt collapse of the polymer chain takes place. Fig. 3 (a) shows the expansion parameter 
$\alpha$ as a function cosolvent concentration for different values of $\tilde{w}_{pc}$. 
At values of bulk cosolvent concentration at which the collapse occurs 
there is also a jump of cosolvent concentration in the gyration volume to very dense packing (fig. 3 (b)).
In contrast to the previous case, in this regime the polymer collapse happens 
as a first - order phase transition at which the jump of the cosolvent 
concentration takes place. As mentioned above, this phase transition is due to the competition 
between polymer - cosolvent attraction, which tends to shrink the polymer chain,
and a steric factor of the cosolvent molecules and monomers, which tends to expand it.

It is interesting to consider the dependence of the expansion parameter 
$\alpha$ on the polymer-cosolvent interaction parameter $\tilde{w}_{pc}$.
As shown in fig. 4 this dependence is sufficiently non-monotonic. 
The collapse of polymer chain takes place in the range of negative values of $\tilde{w}_{pc}$. 
At increasing  $\tilde{w}_{pc}$ the expansion parameter towards zero a maximum occurs. 
Further increasing (towards positive values) $\tilde{w}_{pc}$ the expansion 
parameter again monotonically decreases and smoothly approach the globule regime. Such behavior is in agreement 
with results of computer simulations obtained in \cite{Collaps_2013}.

\begin{figure}[h]
\center{\includegraphics[width=0.65\linewidth]{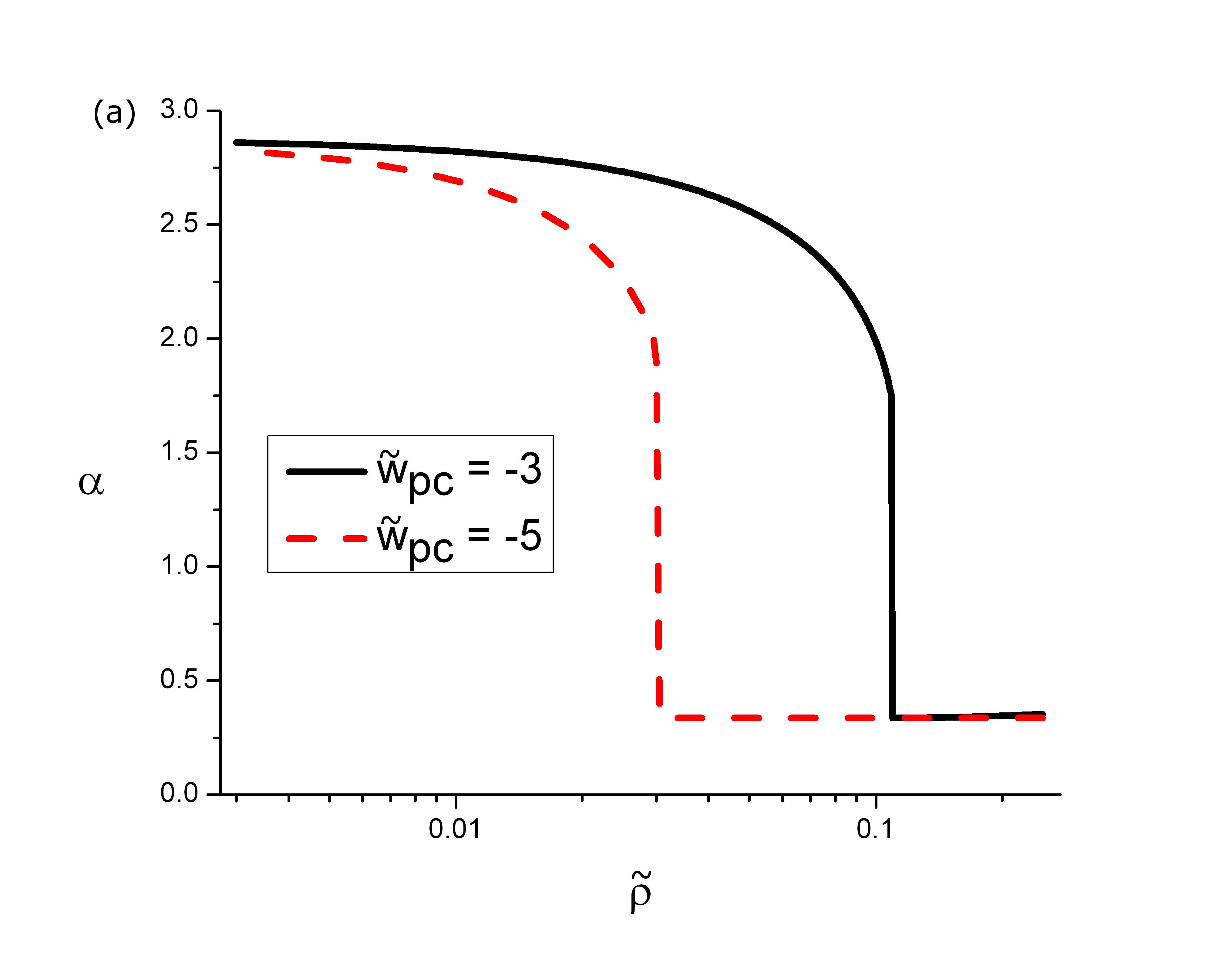}}
\center{\includegraphics[width=0.65\linewidth]{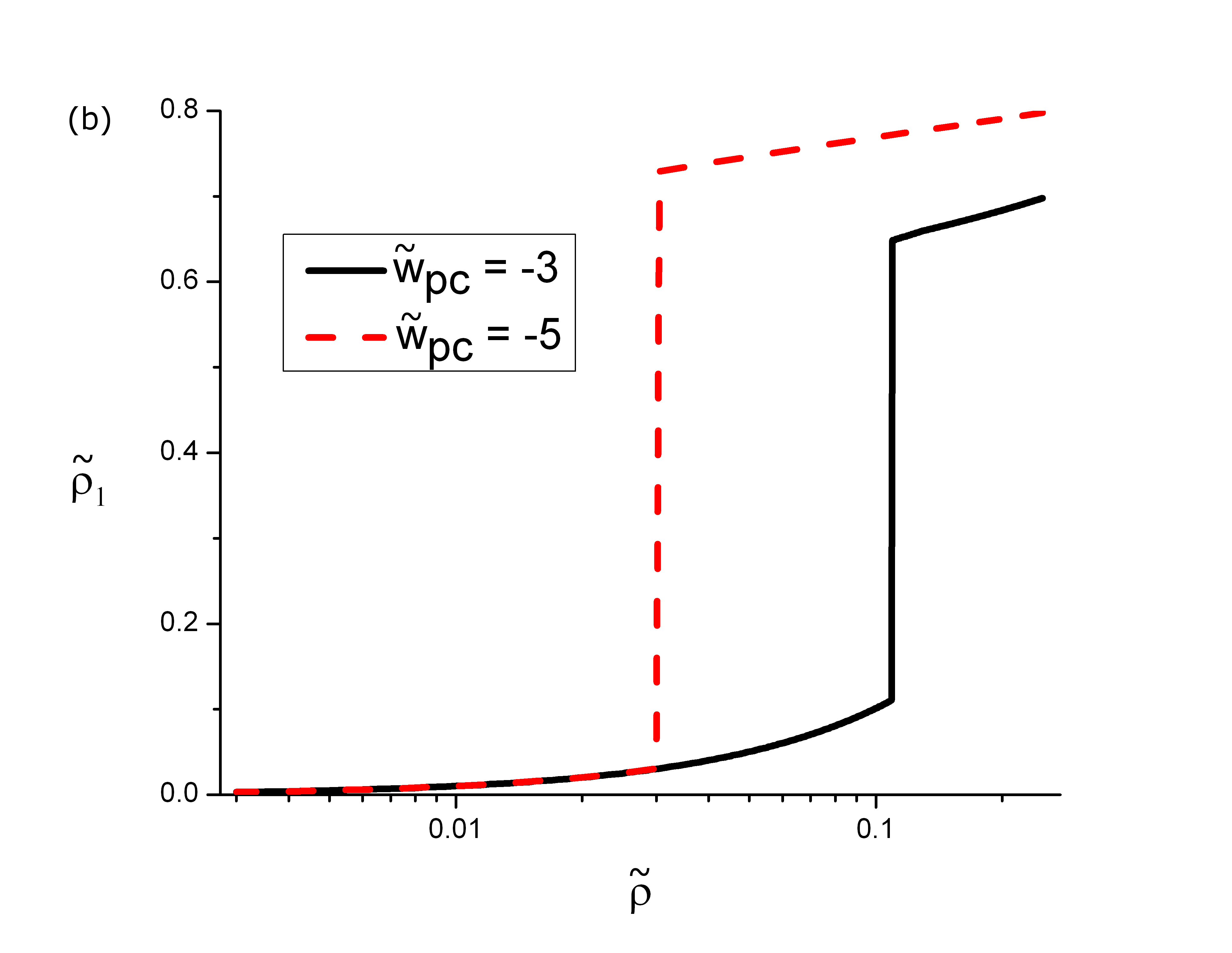}}
\caption{\sl Attractive polymer-cosolvent interaction (e.g. $\tilde{w}_{pc}<0$). 
(a)  The expansion parameter $\alpha$ as a function of the cosolvent bulk concentration $\tilde{\rho}$.  
(b) The cosolvent concentration in the gyration volume $\tilde{\rho}_{1}$ as a function of cosolvent concentration 
in the bulk shown for polymer-cosolvent interaction parameter  $\tilde{w}_{pc}=-3;-5$. 
The bulk cosolvent concentration at which the chain collapse occurs coincides with the jump 
in the cosolvent concentration within the gyration volume. In this regime the polymer collapse 
happens as a first - order phase transition. Values are shown for 
$\tilde{w}_{p}=1$, $N=10^4$, $\tilde{d}_{m}=\tilde{d}_{c}=\tilde{d}_{mc}=1$.}
\label{Fig:3}
\end{figure}

\begin{figure}[h]
\center{\includegraphics[width=0.65\linewidth]{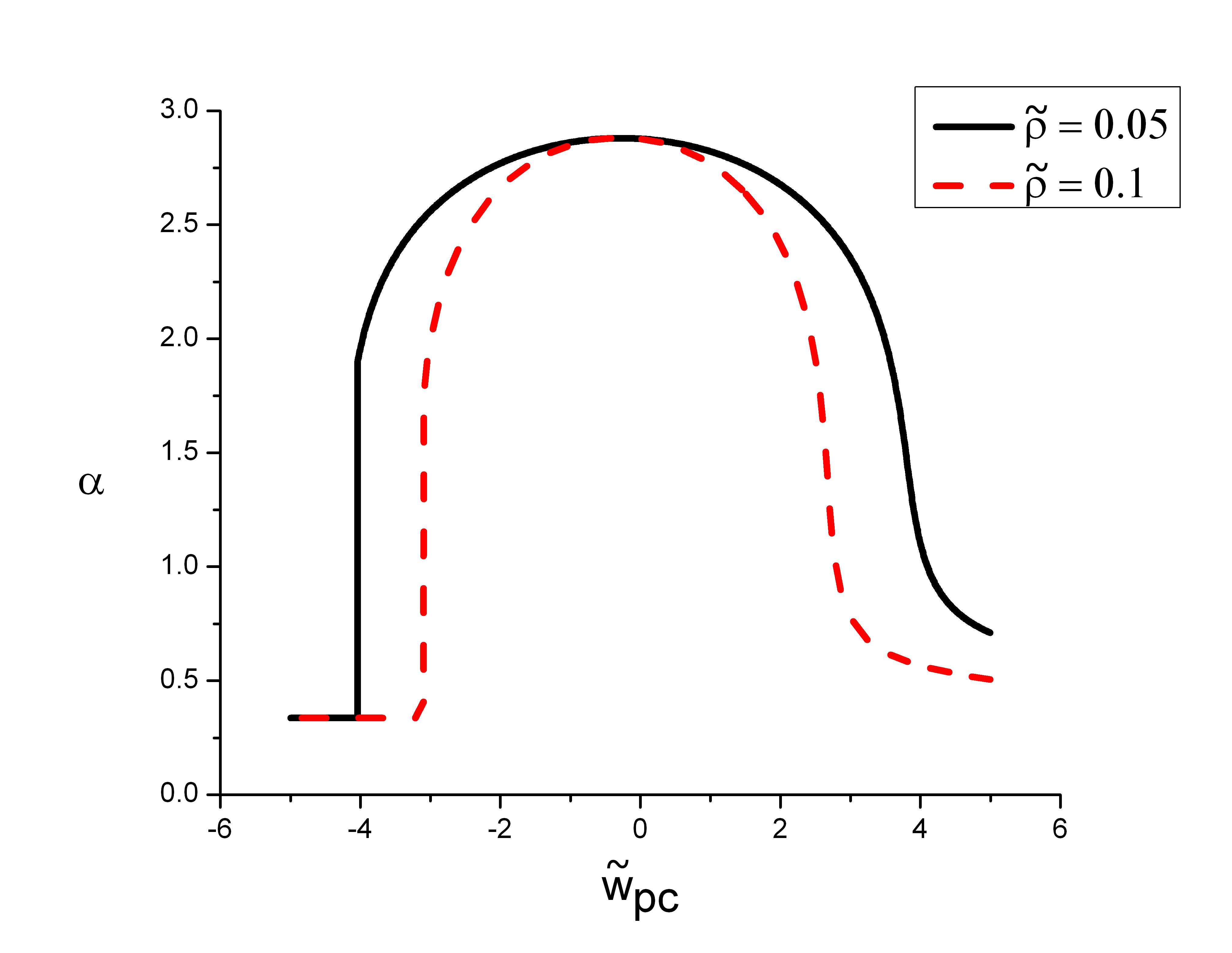}}
\caption{\sl  The expansion parameter $\alpha$ as a function of 
polymer-cosolvent interaction parameter $\tilde{w}_{pc}$ 
at different cosolvent bulk concentrations $\tilde{\rho}$. 
At negative values of $\tilde{w}_{pc}$  the collapse of polymer chain takes place. 
Increasing $\tilde{w}_{pc}$ towards zero a maximum occurs. Further increasing 
(towards positive values) $\tilde{w}_{pc}$ the expansion parameter again 
monotonically decreases and smoothly approach the globule regime. 
Shown here for $\tilde{w}_{p}=1$, $N=10^4$, $\tilde{d}_{m}=\tilde{d}_{c}=\tilde{d}_{mc}=1$.}
\label{Fig:4}
\end{figure}

\section{Summary}
We have outlined a first principles mean-field theory of conformational changes 
of a polymer chain depending on the cosolvent concentration and the type of interactions 
between cosolvent and the polymer. The explicit account of the cosolvent 
leads to the fundamentally new effects, namely polymer chain collapse occurs in two limiting cases.
The first case, when the cosolvent-polymer interaction is a strongly repulsive and, 
in the opposite case, when this interaction is strongly attractive.
In the second case when polymer-cosolvent interaction is attractive the collapse occurs 
as a first-order phase transition, i.e. discontinuous change of the radius 
of gyration and the cosolvent density within the gyration volume. We call these phenomena 
“cosolvent-induced coil-globule transitions”.

The described phenomena may be relevant for applications, where the mixed-solvent polymer solutions are used. 
In particular where the solvent conditions are used to induce a change in polymer conformation the described 
mechanism might offer an additional routine to exert control on the polymer conformational transition.
In particular, the present theory may be useful for description of stabilization of 
proteins in aqueous solution by adding osmolytes, such as trimethylamine N-oxide (TMAO), glycine, 
betaine, glycerol, and sugars \cite{Cosolvent_2013}. We would like to stress that the 
effect of the polymer collapse in the case of strong attraction between monomers 
and cosolvent molecules is very similar to the collapse of polypeptide PNIPAM 
due to cooperative generating of hydrogen bonding of urea with the polypeptide backbone \cite{Cremer_review}. 
However, these speculations require more detailed investigations involving the 
computer simulations and comparisons with real experimental data, that is a subject of the forthcoming publications.

Instructively to provide an estimation in physical units of values of cosolvent 
concentration in the bulk $\rho$ and second virial coefficient of interaction 
monomer-cosolvent $w_{pc}$ at which the coil-globule transitions can take place. 
We consider the case when the value of Kuhn's length of the segment is approximately 
equal to $b\approx d_{m}\approx d_{c} \approx d_{mc}\approx 0.4~nm$, 
that is quite reasonable for real polymer solutions \cite{Collaps_2013,Cremer_review}. 
In the case when interaction of monomer-cosolvent is pure repulsive 
one can get the following estimates: $\rho \sim 10~ M$, 
$w_{pc}\sim 10^2$ $A^3$. In the case of attractive interaction we obtain the estimates:
$\rho\sim 1~M$ and $|w_{pc}|\sim 10^{2}$ $A^{3}$. These estimates show that the reported coil-globule transitions 
may be observed under ambient conditions in reality.

The present theory, however has natural limitations. Firstly, it can not describe a dilute 
polyelectrolyte solutions, where many-body effects due to long-range electrostatic 
interactions play crucial role. Such a first-principals theories, 
which can describe the collapse of the charged polymer chain 
in dilute polyelectrolyte solution, has been recently developed in works \cite{Brilliantov_collaps,Pincus}.     
Moreover, our theory does not take into account a possibility the formation of chemical bonds 
between polymer and cosolvent molecules. Apart from the above limitations 
our theory does not explicitly account for specific interactions, 
such as hydrogen bonds formation. We note that work in this direction 
has been recently published  \cite{Podgornik}, where 
the problem of the helix-coil transition in explicit solvent has been addressed analytically. Employing a
spin-based models the influence of the hydrogen bonds formation on 
the helix-coil transition has been investigated. 
It would be interesting  to investigate how the chemical bond formation 
and explicit accounting specific interactions influence 
on the coil-globule transition in dilute polymer solutions.  
We believe, that these problems also can be a subject worthwhile for forthcoming publications.

\section{Acknowledgments}
The research leading to these results has received funding from the European Union's Seventh Framework Programme (FP7/2007-2013) under grant 
agreement №//247500 with //project acronym "Biosol".

We are grateful to prof. N.V. Brilliantov for his remarks which allowed to do a more rigorous theory.

We thank anonymous referees for their valuable suggestions and comments.

\section{Appendix: derivation of formula (\ref{PartSum:CoSolv})}
To address a derivation of the expression (\ref{PartSum:CoSolv}) 
for the partition function of the cosolvent we start from the canonical partition 
function of the solution, which can be written as follows
\begin{equation}
Z(R_{g})=\int d\Gamma_{p}(R_{g})\int d\Gamma_{c}\exp\left[-\beta H_{p}-\beta H_{c}-\beta H_{pc}\right],
\end{equation}
where the symbol $\int d\Gamma(R_{g})(..)$ denotes the integration over microstates 
of the polymer chain performed at a fixed radius of gyration;
the symbol $\int d\Gamma_{c}(..)=\frac{1}{N_{c}!}\int\limits_{V} d\vec{r}_{1}..\int\limits_{V} d\vec{r}_{N_{c}}(..)$ 
denotes the integration over cosolvent molecules' coordinates; $V$ is a volume of the system;
\begin{equation}
\beta H_{p}=\frac{w_{p}}{2}\int\limits_{0}^{N}ds_{1}\int\limits_{0}^{N}ds_{2}\delta\left(\vec{r}(s_{1})-\vec{r}(s_{2})\right)+\frac{\beta}{2}\int\limits_{0}^{N}ds_{1}\int\limits_{0}^{N}ds_{2} V_{hc}^{(m)}(\vec{r}(s_{1})-\vec{r}(s_{2})) 
\end{equation}
is the Hamiltonian of  the monomer-monomer interaction; 
$w_{p}$ is a second virial coefficient for the monomer-monomer interaction and
\begin{equation}
V_{hc}^{(m)}(\vec{r})=\Biggl\{
\begin{aligned}
\infty,\quad&|\vec{r}|\leq d_{m}\,\\
0,\quad& |\vec{r}|> d_{m}
\end{aligned}
\end{equation}
is a hard-core potential of interaction monomer-monomer;
$d_{m}$ is a diameter of monomers;
\begin{equation}
\beta H_{pc}=w_{pc}\int\limits_{0}^{N}ds\sum\limits_{j=1}^{N_{c}}\delta\left(\vec{r}(s)-\vec{r}_{j}\right)+\beta \int\limits_{0}^{N}ds\sum\limits_{j=1}^{N_{c}}V_{hc}^{(mc)}(\vec{r}(s)-\vec{r}_{j})
\end{equation}
\begin{equation}
=w_{pc}\int\limits_{V} d\vec{x}\hat{\rho}_{c}(\vec{x})\hat{\rho}_{p}(\vec{x})+\beta H_{pc}^{(hc)}
\end{equation}
is the Hamiltonian of the polymer-cosolvent interaction;
$\hat{\rho}_{c}(\vec{x})=\sum\limits_{i=1}^{N_{c}}\delta\left(\vec{x}-\vec{r}_{i}\right)$ and 
$\hat{\rho}_{p}(\vec{x})=\int\limits_{0}^{N} ds\delta(\vec{x}-\vec{r}(s))$
are the local densities of the cosolvent molecules and monomers, respectively;  
$w_{pc}$ is the second virial coefficient for the polymer-cosolvent interaction,
\begin{equation}
V_{hc}^{(mc)}(\vec{r})=\Biggl\{
\begin{aligned}
\infty,\quad&|\vec{r}|\leq d_{mc}\,\\
0,\quad& |\vec{r}|> d_{mc}
\end{aligned}
\end{equation}
is a hard-core potential of the monomer-cosolvent interaction ;
\begin{equation}
\beta H_{c}=\frac{\beta}{2}\sum\limits_{j\neq{i}}V_{hc}^{(c)}(\vec{r}_{i}-\vec{r}_{j})
\end{equation}
is the Hamiltonian of the cosolvent-cosolvent interaction; 
\begin{equation}
 V_{hc}^{(c)}(\vec{r})=\Biggl\{
\begin{aligned}
\infty,\quad&|\vec{r}|\leq d_{c}\,\\
0,\quad& |\vec{r}|> d_{c}
\end{aligned}
\end{equation}
is the hard-core potential ($d_{c}$ is a cosolvent molecule diameter). 
Thus we describe the cosolvent-cosolvent interaction as an excluded volume interaction.

Making the following identity transformation
\begin{equation}
Z(R_{g})=\int d\Gamma_{p}(R_{g})e^{-\beta H_{p}}\int d\Gamma_{c}e^{-\beta H_{c}-\beta H_{pc}}=Z_{p}(R_{g})\int d\Gamma_{c}e^{-\beta H_{c}}\left<e^{-\beta H_{pc}}\right>_{p},
\end{equation}
where 
\begin{equation}
Z_{p}(R_{g})=\int d\Gamma_{p} (R_{g})e^{-\beta H_{p}}
\end{equation}
is the polymer partition function; the symbol $\left<(..)\right>_{p}=\frac{1}{Z_{p}(R_{g})}\int d\Gamma(R_{g})e^{-\beta H_{p}}(..)$ 
denotes  averaging over polymer microstates with a fixed radius of gyration. 
To take into account the polymer-cosolvent hard-sphere effect we will assume 
that the contribution of the polymer-cosolvent hard-core interaction in the total Hamiltonian 
just leads to a renormalization of the cosolvent configurational space. This amounts to 
\begin{equation}
\label{renorm_conf}
\int d\Gamma_{c} e^{-\beta H_{pc}^{(hc)}}(..)\rightarrow \int d\Gamma_{c}^{\prime}(..),
\end{equation}
where $\int d\Gamma_{c}^{\prime}(..)=\frac{1}{N_{c}!}\int\limits_{V-Nv_{mc}}d\vec{r}_{1}..\int\limits_{V-Nv_{mc}}d\vec{r}_{N_{c}}(..)$.
It should be noted that the approximation (\ref{renorm_conf}) is the simplest method to account for  the 
so-called depletion forces \cite{Barrat_Hansen} which may lead to an additional repulsion of the cosolvent molecules 
from the gyration volume due to the presence of monomers. 
Keeping in mind the "mean-field" assumption (\ref{renorm_conf}) we arrive at
\begin{equation}
Z(R_{g})=Z_{p}(R_{g})\int d\Gamma_{c}^{\prime}e^{-\beta H_{c}}\left<e^{-w_{pc} \int\limits_{V} d\vec{x}\hat{\rho}_{c}(\vec{x})\hat{\rho}_{p}(\vec{x})}\right>_{p}.
\end{equation}

Truncating the cumulant expansion \cite{Kubo} at the first order we obtain 
\begin{equation}
\left<e^{-w_{pc}\int\limits_{V}\hat{\rho}_{c}(\vec{x})\hat{\rho}_{p}(\vec{x})}\right>_{p}\approx e^{-w_{pc}\int\limits_{V} d\vec{x}\hat{\rho}_{c}(\vec{x})\left<\hat{\rho}_{p}(\vec{x})\right>_{p}}.
\end{equation}
Thus one can get
\begin{equation}
w_{pc}\int\limits_{V} d\vec{x}\hat{\rho}_{c}(\vec{x})\left<\hat{\rho}_{p}(\vec{x})\right>_{p}\simeq \frac{w_{pc}N}{V_{g}}\int\limits_{V_{g}}d\vec{x}\hat{\rho}_{c}(\vec{x}),
\end{equation}
where the approximation
\begin{equation}
 \left<\hat {\rho}(\vec{x})\right>_{p}\simeq \Biggl\{
\begin{aligned}
\frac{N}{V_{g}},\quad&|\vec{x}|\leq R_{g}\,\\
0,\quad& |\vec{x}|> R_{g},
\end{aligned}
\end{equation}
has been introduced.
Therefore, we obtain the following expression for the partition function of the solution
\begin{equation}
Z(R_{g})=Z_{p}(R_{g})Z_{c}(R_{g}),
\end{equation}
where $Z_{c}(R_{g})$ has a form
\begin{equation}
Z_{c}(R_{g})=\int d\Gamma_{c}^{\prime}e^{-\beta H_{c}-\frac{w_{pc}N}{V_{g}}\int\limits_{V_{g}}d\vec{x}\hat{\rho}_{c}(\vec{x})}.
\end{equation}
Rewriting the last expression in the form
\begin{equation}
Z_{c}(R_{g})=\frac{1}{N_{c}!}\sum\limits_{n=0}^{N_{c}}Z_{c}(R_{g},n),
\end{equation}
where
\begin{equation}
Z_{c}(R_{g},n)=\frac{N_{c}!}{(N_{c}-n)!n!}\int\limits_{V_{g}-Nv_{mc}}d\vec{x}_{1}..\int\limits_{V_{g}-Nv_{mc}}d\vec{x}_{n}\int\limits_{V-V_{g}}d\vec{y}_{1}..\int\limits_{V-V_{g}}d\vec{y}_{N_{c}-n}
e^{-\beta H_{c}-\beta H_{int}}
\end{equation}
is the cosolvent partition function with a fixed number $n$ of cosolvent molecules in the gyration volume; 
$\beta H_{int}=\frac{w_{pc}N}{V_{g}}\int\limits_{V_{g}}d\vec{x}\hat{\rho}_{c}(\vec{x})$ is 
an effective Hamiltonian which describes the monomer-cosolvent interactions.
To evaluate $Z_{c}(R_{g},n)$ we introduce the approximation $H_{c}=H_{c}^{(N_{c})}\approx H_{c}^{(n)}+H_{c}^{(N_{c}-n)}$
which is accurate for sufficiently large  gyration volumes. This leads to 
\begin{equation}
Z_{c}(R_{g},n)\simeq\frac{N_{c}!}{(N_{c}-n)!n!}Z_{c}^{(N_{c}-n)}Z_{c}^{(n)}
\end{equation}
where
\begin{equation}
Z_{c}^{(N_{c}-n)}=\int\limits_{V-V_{g}}d\vec{y}_{1}..\int\limits_{V-V_{g}}d\vec{y}_{N_{c}-n}e^{-\beta H_{c}^{(N_{c}-n)}},
\end{equation}
\begin{equation}
Z_{c}^{(n)}=\int\limits_{V_{g}-Nv_{mc}}d\vec{x}_{1}..\int\limits_{V_{g}-Nv_{mc}}d\vec{x}_{n}e^{-\beta H_{c}^{(n)}-\beta H_{int}}=
Q_{c}^{(n)}\left<e^{-\beta H_{int}}\right>_{c},
\end{equation}
$Q_{c}^{(n)}=\int\limits_{V_{g}-Nv_{mc}}d\vec{x}_{1}..\int\limits_{V_{g}-Nv_{mc}}d\vec{x}_{n}e^{-\beta H_{c}^{(n)}}$ is a configurational integral of the cosolvent 
in gyration volume; the symbol $\left<(..)\right>_{c}=\frac{1}{Q_{c}^{(n)}}\int\limits_{V_{g}-Nv_{mc}}d\vec{x}_{1}..\int\limits_{V_{g}-Nv_{mc}}d\vec{x}_{n}e^{-\beta H_{c}^{(n)}}(..)$ means an averaging
over cosolvent microstates in gyration volume.
Now we have to evaluate the cosolvent partition function $Z_{c}^{(n)}$ in the gyration volume. 

Truncating the cumulant expansion at the second order we obtain:
\begin{equation}
Z_{c}^{(n)}=Q_{c}^{(n)}\exp\left[-\beta \left<H_{int}\right>_{c}+\frac{\beta^2}{2}\left(\left<H_{int}^2\right>_{c}-\left<H_{int}\right>_{c}^2\right)\right].
\end{equation} 
Then we obtain:
\begin{equation}
\beta \left<H_{int}\right>_{c}-\frac{\beta^2}{2}\left(\left<H_{int}^2\right>_{c}-\left<H_{int}\right>_{c}^{2}\right)=\frac{w_{pc}N}{V_{g}}\int\limits_{V_{g}}d\vec{x}\left<\hat{\rho}_{c}(\vec{x})\right>_{c}-\frac{w_{pc}^2N^2}{2V_{g}^2}\int\limits_{V_{g}}d\vec{x}\int\limits_{V_{g}}d\vec{y}\left<\delta\hat{\rho}_{c}(\vec{x})\delta\hat{\rho}_{c}(\vec{y})\right>_{c}.
\end{equation}
Using a local approximation for the correlation function of the cosolvent density fluctuations in the gyration volume
\begin{equation}
\left<\delta\hat{\rho}_{c}(\vec{x})\delta\hat{\rho}_{c}(\vec{y})\right>_{c}=\frac{n}{V_{g}}\frac{\chi_{T}}{\chi_{T}^{id}}\delta(\vec{x}-\vec{y}),
\end{equation}
and keeping in mind that 
$\left<\hat{\rho}_{c}(\vec{x})\right>_{c}=\frac{n}{V_{g}-Nv_{mc}}$ we arrive at
\begin{equation}
Z_{c}^{(n)}=Q_{c}^{(n)}\exp\left[-\frac{w_{pc}nN}{V_{g}-Nv_{mc}}+\frac{w_{pc}^2N^2n\chi_{T}}{2V_{g}^2\chi_{T}^{id}}\right],
\end{equation}
where $\chi_{T}$ is an isothermal compressibility of cosolvent in the gyration volume and $\chi_{T}^{id}$ 
is an isothermal compressibility of the ideal gas.

Applying the mean-field approximation to $Z_{c}^{(N_{c}-n)}$ and $Q_{c}^{(n)}$ we finally arrive at 
\begin{equation}
\label{Zc}
Z_{c}(R_{g})=\sum\limits_{n=0}^{N_{c}}\frac{(V_{g}-nv_{c}-Nv_{mc})^{n}(V-V_{g}-(N_{c}-n)v_{c})^{N_{c}-n}}{(N_{c}-n)!n!}e^{-\beta \Delta F_{int}(n)},
\end{equation}
where 
\begin{equation}
\label{Fluct_free}
\beta \Delta F_{int}(n)=\frac{w_{pc}Nn}{V_{g}-Nv_{mc}}-\frac{w_{pc}^2N^2n\chi_{T}}{2V_{g}^2\chi_{T}^{id}},
\end{equation}
$v_{c}=\frac{\pi d_{c}^3}{6}$ is an excluded volume of the cosolvent molecules. The first term on the right hand side 
of (\ref{Fluct_free}) determines the contribution of 
monomer-cosolvent interaction in the framework the mean-field approximation.
The second term is related to an effective  monomer-monomer interaction due to the cosolvent concentration fluctuations.  
In the thermodynamic limit ($N_{c}\rightarrow \infty$) in the sum (\ref{Zc}) 
only the highest order term giving the main contribution is relevant. 
This term corresponds to the number $n=N_{1}$ which can be obtained from the extremum condition
\begin{equation}
\frac{\partial}{\partial{n}}\ln\left(\frac{(V_{g}-nv_{c}-Nv_{mc})^{n}(V-V_{g}-(N_{c}-n)v_{c})^{N_{c}-n}}{(N_{c}-n)!n!}e^{-\beta\Delta F_{int}(n)}\right)=0.
\end{equation}
Therefore we arrive at the expression for the cosolvent partition function 
in the framework of the mean-field approximation which already has been used in the main text
\begin{equation}
Z_{c}(R_{g})\simeq \frac{(V_{g}-N_{1}v_{c}-Nv_{mc})^{N_{1}}(V-V_{g}-(N_{c}-N_{1})v_{c})^{N_{c}-N_{1}}}{(N_{c}-N_{1})!N_{1}!}e^{-\beta \Delta F_{int}(N_{1})}.
\end{equation}

\end{document}